\documentclass[manuscript, acmsmall, screen, dvipsnames]{acmart}

\usepackage[utf8]{inputenc}
\usepackage[T1]{fontenc}

\usepackage{mathtools}
\usepackage{booktabs}
\usepackage{tabularx}
\usepackage{listings}
\usepackage{amsmath}
\usepackage{amsfonts}
\usepackage{pgfplotstable}

\usepackage{tabularx}
\usepackage{array}
\usepackage{caption}
\newcolumntype{L}{>{\raggedright\arraybackslash}X}
\newcolumntype{Y}{>{\raggedleft\arraybackslash}X}

\colorlet{punct}{red!60!black}
\definecolor{background}{HTML}{FFFFFF}
\definecolor{delim}{RGB}{20,105,176}
\colorlet{numb}{magenta!60!black}

\lstdefinelanguage{json}{
    basicstyle=\normalfont\ttfamily,
    showstringspaces=false,
    breaklines=true,
    backgroundcolor=\color{background},
    literate=
     *{0}{{{\color{numb}0}}}{1}
      {1}{{{\color{numb}1}}}{1}
      {2}{{{\color{numb}2}}}{1}
      {3}{{{\color{numb}3}}}{1}
      {4}{{{\color{numb}4}}}{1}
      {5}{{{\color{numb}5}}}{1}
      {6}{{{\color{numb}6}}}{1}
      {7}{{{\color{numb}7}}}{1}
      {8}{{{\color{numb}8}}}{1}
      {9}{{{\color{numb}9}}}{1}
      {:}{{{\color{punct}{:}}}}{1}
      {,}{{{\color{punct}{,}}}}{1}
      {\{}{{{\color{delim}{\{}}}}{1}
      {\}}{{{\color{delim}{\}}}}}{1}
      {[}{{{\color{delim}{[}}}}{1}
      {]}{{{\color{delim}{]}}}}{1},
}

\DeclareMathOperator{\rn}{RN}
\DeclareMathOperator{\rz}{RZ}
\DeclareMathOperator{\rd}{RD}
\DeclareMathOperator{\ru}{RU}
\DeclareMathOperator*{\argmax}{arg\,max}
\DeclareMathOperator{\sign}{sign}

\def\fmin{f_\mathrm{min}}
\def\smin{s_\mathrm{min}}
\def\fmax{f_\mathrm{max}}
\def\emin{e_\mathrm{min}}
\def\emax{e_\mathrm{max}}

\newcommand{\R}{\ensuremath{\mathbb}{R}}
\newcommand{\F}{\ensuremath{\mathbb}{F}\langle \emin, \emax, p\rangle}
\newcommand{\fp}{floating-point}
\newcommand{\ulp}[1]{\ensuremath{\mathrm{ulp}(#1)}}
\newcommand{\err}[1]{\ensuremath{E(#1)}}

\theoremstyle{thmstyletwo}%
\numberwithin{equation}{section}

\setcopyright{none}
\renewcommand\footnotetextcopyrightpermission[1]{}

\begin{document}

\title[Accuracy of Mathematical Functions in Julia]{Accuracy of Mathematical Functions in Julia}

\author{Mantas Mikaitis}
\email{m.mikaitis@leeds.ac.uk}
\orcid{0000-0001-8706-1436}
\affiliation{%
  \institution{School of Computer Science, University of Leeds}
  \city{Leeds}
  \country{UK}
}
\author{Mos\`{e} Giordano}
\email{m.giordano@ucl.ac.uk}
\orcid{0000-0002-7218-2873}
\affiliation{%
  \institution{Advanced Research Computing Centre, University College London}
  \city{London}
  \country{UK}
}
\author{Tejaswa Rizyal}
\email{tejaswarizyal4@gmail.com}
\affiliation{%
  \institution{School of Computer Science, University of Leeds}
  \city{Leeds}
  \country{UK}
}


\begin{abstract}
  Basic computer arithmetic operations, such as $+$, $\times$, or $\div$ are correctly rounded, whilst mathematical functions such as $e^x$, $\ln(x)$, or $\sin(x)$ in general are not, meaning that separate implementations may provide different results when presented with an exact same input, and that their accuracy may differ.
  We present a methodology and a software tool that is suited for exhaustive and non-exhaustive testing of mathematical functions available as part of the Julia programming language, in various floating-point formats.
  The software tool is useful to the users of Julia, to quantise the level of accuracy of the mathematical functions and interpret possible effects of errors on their scientific computation codes that depend on these functions.
  It is also useful to the developers and maintainers of the functions in Julia Base, to test the modifications to existing functions and to test the accuracy of new functions.
  The software (a test bench) is designed to be easy to set up for running the accuracy tests in automatic regression testing.
  Our focus is to provide software that is user friendly and allows to avoid the need for specialised knowledge of floating-point arithmetic or the workings of mathematical functions; users only need to supply a list of formats, choose the rounding modes, and specify the input space search strategies based on how long they can afford the testing to run.
  We have utilized the test bench to determine the errors of a subset of mathematical functions in Julia 1.12.7, for binary16, binary32, and binary64 IEEE 754 floating-point formats, and found $0.49$ to $0.51$ULPs in binary16, and $0.49$ to $2.4$ULPs of error in binary32 and binary64.
  The functions that are correctly rounded (error below $0.5$ULP) in binary 16 and binary32 are sqrt and cbrt, and in sqrt binary64.
  The following functions are correctly rounded only for binary16: asin, atanh, cospi, log2, sinh, sinpi, tanh.
\end{abstract}


\received{22 December 2025}
\received[revised]{22 December 2025}

\maketitle

\begin{center}
  {\bf Dedicated to the memory of Nicholas J. Higham}
\end{center}

\section{Introduction}

Take $f$ to be a \emph{transcendental function} over $x \in \R$.
On a computer we often approximate $f(x)$ for some $x$ in a finite subset of the reals, such as floating-point numbers of certain precision.
In general, $f$ cannot be computed exactly: for example, the exponential function $f(x)=e^x$ has a solution
\begin{equation}\label{eq:exp-series}
  e^x = \sum_{k=0}^{\infty}\frac{x^k}{k!} = 1 + x + \frac{x^2}{2} + \frac{x^3}{6}+\frac{x^4}{24} + \cdots.
\end{equation}
While this series is in general not the best way to approximate the exponential function, with other approaches such as piece-wise polynomials and table-based algorithms in a wider use~\citep{mull16}, the following applies in general.

There are multiple sources of error in approximating $e^x$.
The first error source is in truncating the infinite series; the second error is induced in evaluating it on a computer with limited precision arithmetic; finally, the third error comes from rounding the approximation to some floating-point precision.
Given a target accuracy for the approximation of the exponential function, how many steps of \eqref{eq:exp-series} do we need to evaluate?
The answer to that question affects the internal precision at which the approximation needs to be computed before rounding to some final target precision.
In some cases, when the approximation is very close to the half-way value (a \emph{breakpoint}~\citep[Sec.~12.3]{mull16}) between two floating-point values that may be returned by rounding, the extra precision required to determine on which side of the breakpoint we should be, when rounding to nearest, may need to be very large~\citep[Sec.~12.3]{mull16}---the problem of determining how big the internal precision needs to be in order for the final rounding to be correct is termed \emph{table maker's dilemma}~(TDM).
Each rounding mode has different breakpoints and for directed rounding the breakpoints are the floating-point numbers themselves.
In general, table maker's dilemma is not solved and therefore the accuracy of mathematical functions is not guaranteed to be optimal.
Even when algorithms for finding hardest-to-round cases for a generalised floating-point format and all functions of interest are discovered, the implementations may be expensive and the vendors may choose not to put resources into handling those special cases, or it will take time for the theoretical research to be applied in practice.
When an implementation does not provide guarantees for the optimal accuracy of some functions in some floating-point format, testing is required to determine the maximum errors.
This may be of interest to a wide set of people,  some of who may not want to program at floating-point bit-level representation---users of the functions in scientific applications, Julia developers, or computer arithmetic researchers~\citep{gimz25}---therefore the testing software should be easy to set-up and the testing should be straightforward to rerun automatically for when a particular mathematical function is modified by the developer.

An illustrative example of table-maker's dilemma can be shown in a 16-bit floating-point format called \emph{bfloat16} (see Section~\ref{sec:fp}) which has 8 binary digits of precision.
Take $\frac{47}{32}=1.46875$, a quantity that is exactly representable in bfloat16.
For the purposes of demonstration we will assume that range reducation has already been done---in reality $1.46875$ would be reduced to something much smaller before evaluating the approximation of the exponential function.
Expressed in binary,
$$
e^{1.46875}=\fbox{100.01011}\:00000000011\cdots
$$ 
which was discovered by Lef\`{e}vre~and~Muller using an exhaustive search~\citep{lemu03}.
The boxed part of the number is the part that will be retained, with a potential rounding of the digits that are outside the box.
Assume that we choose to round the number toward zero and denote the rounding operation as $\text{RZ}()$ (see Table~\ref{table:round}).
Then $\text{RZ}\Big(\:\fbox{100.01011}\:00000000011\cdots\Big)=100.01011$.
Notice that the exact value of this exponential is very close to a number representable in bfloat16 since it is followed by 9 digits of zero before 1's start to appear.
Because the exact result is close to a rounding breakpoint, a poor approximation of $e^{1.46875}$ may end up below the breakpoint.
Indeed, computing the exponential $e^{1.46875}$ with the series in Equation~\ref{eq:exp-series} truncated to $k=8$ we get
$$
\fbox{100.01010}1111111110010\cdots
$$
and rounding to zero would yield an incorrect value in the context of \emph{correct rounding} of IEEE 754~\cite{ieee19} and the definition of $\rz$ in Table~\ref{table:round}.
However, running the series more accurately, to $k=9$, we obtain
$$
\fbox{100.01011}\: 00000000010\cdots
$$
and a correct round-to-zero result would be computed.
The input $x=1.46875$ to the exponential function with the 8-digit arithmetic is called \emph{a difficult case} and the cases where the approximation needs to be the most accurate are called \emph{hardest-to-round points}~\citep[Sec.~12.3]{mull16}.
With the hardest-to-round input the approximation of the exponential function requires the largest accuracy for the 8-digit result to end up correct, compared with all the possible 8-digit inputs.
Having knowledge of the hardest-to-round case allows programmers to target the accuracy of the approximation of a function in order to obtain the correctly rounded result.
In general hardest-to-round points for all functions and input/output precisions are not known, but for some commonly used floating-point formats these points have been discovered~\citep{bhmz25}.

The problem outlined above is not present with \emph{algebraic functions} involving a finite number of terms with algebraic operations, such as $\sqrt{x}$, $x+y$, $x\times y$ or $1/x$.
The 2019 revision of the IEEE 754~\citep{ieee19} floating-point standard \emph{requires} implementations to provide some correctly rounded operations such as $+$, $-$, $\times$, $\sqrt{\vphantom{x}}$, and $\div$, but only \emph{recommends} to provide correctly rounded transcendental functions such as the exponential and logarithm.
IEEE 754 lists the following ``\emph{additional mathematical operations}'' (some specific cases of the ones listed are omitted) and underlined are the ones that we test in this paper.

\begin{center}

\underline{$e^x$}, \; $e^x-1$ (expm1), \; \underline{$2^x$}, \; $2^x-1$ (exp2m1), \; \underline{$10^x$}, \; $10^x-1$ (exp10m1), \;

\underline{$\log_e(x)$}, \; \underline{$\log_2(x)$}, \; \underline{$\log_{10}(x)$}, \; \underline{$\log_e(1+x)$} (logp1), \; $\log_2(1+x)$ (log2p1), \; $\log_{10}(1+x)$ (log10p1), \;

$\sqrt{x^2+y^2}$ (hypot), \; $1/\sqrt{x}$, \; $(1+x)^n$ (compound), \;
$x^{1/n}$ (rootn), \;
$x^n$, \; $x^y$, \;

\underline{$\sin(x)$}, \; \underline{$\cos(x)$}, \; \underline{$\tan(x)$}, \; \underline{$\sin(x\pi)$} (sinpi), \; \underline{$\cos(x\pi)$} (cospi), \; \underline{$\tan(x\pi)$} (tanpi), \;

\underline{$\mathrm{asin}(x)$}, \; \underline{$\mathrm{acos}(x)$}, \; \underline{$\mathrm{atan}(x)$}, \;
\underline{$\mathrm{sinh}(x)$}, \;
\underline{$\mathrm{cosh}(x)$}, \;
\underline{$\mathrm{tanh}(x)$}, \;
\underline{$\mathrm{asinh}(x)$}, \;
\underline{$\mathrm{acosh}(x)$}, \;
\underline{$\mathrm{atanh}(x)$}.

\end{center}

Here $x$ and $y$ are in the extended reals~\cite{ieee19}, and $n$ is an integer.

We additionally test $\sqrt{x}$ and $\sqrt[3]{x}$ (cbrt), the former of which is required by IEEE 754, and the latter is not listed in IEEE 754.
The names of lesser-known functions are listed in brackets, with universally known ones like exp and log, omitted.

In summary, in this work we report the accuracy of the underlined functions, in Julia programming language.
We do not test just whether they are correctly rounded and report yes or no for each---we instead report worst-case errors in binary16 and binary32 formats, and report lower bounds on the worst case errors for binary64.

\section{Background}

\subsection{Julia programming language}

Julia~\citep{beks17} is a high-level dynamic programming language, focused on numerical computing and high performance, but now widely used for general purposes, such as data analysis, machine learning, or visualisation.
For example, it is used for machine learning~\cite{gmpc20}, simulation of open quantum systems~\cite{kpor18}, and high-energy physics~\cite{sbgh25}.
Similar to MATLAB, Octave and Python, it is dynamically typed, meaning that users do not need to declare types of variables, such as the floating-point format; annotating Julia programs with types is optional~\citep[Sec.~3.4]{beks17}.
At the time of writing, Julia has been downloaded over 100 million times\footnote{\url{https://julialang.org}} and the TIOBE Programming Community index\footnote{\url{https://www.tiobe.com/tiobe-index/julia/}} listed it at top 26 in terms of popularity in August 2026, with the highest ranking being top 20 in August 2023.

Many mathematical functions in Julia are implemented directly\footnote{\url{https://github.com/JuliaLang/julia/tree/master/base/special}} instead of being based on one of the mathematical function libraries~\citep{gimz25}.
Some functions on some specific system configurations may rely on OpenLibm (\url{https://github.com/JuliaMath/openlibm/}).
Limited amount of testing of these functions is provided within the Julia source code.\footnote{\url{https://github.com/JuliaLang/julia/blob/master/test/math.jl}}
Julia also provides fast version computer arithmetic through the @fastmath mode,\footnote{\url{https://github.com/JuliaLang/julia/blob/master/base/fastmath.jl}} which favours performance rather than compliance with the IEEE 754 standard.

The default fixed precision floating-point formats in Julia are binary16, binary32, and binary64 implemented through \texttt{Float16}, \texttt{Float32} and \texttt{Float64} types.
In order to measure the accuracy of Julia's approximations to mathematical functions we will need approximations of the same functions in higher precision than these formats.
For arbitrary precision computation Julia provides an interface\footnote{\url{https://docs.julialang.org/en/v1/manual/integers-and-floating-point-numbers}} to the MPFR library~\citep{fhlp07} and an associated type \texttt{BigFloat}.
MPFR provides correctly rounded elementary functions for a given precision and rounding mode; see~\cite{mpfr-tech-rep} for the list of functions and algorithms used to compute them.
Since MPFR provides implementations of various mathematical functions in arbitrary precision, this is a good option for us.
Additionally, interfaces to  Arb C library for arbitrary precision interval arithmetic~\cite{fred17}, such as \texttt{Arblib.jl}\footnote{\url{https://github.com/kalmarek/Arblib.jl}} and \texttt{ArbNumerics.jl}\footnote{\url{https://juliapackages.com/p/arbnumerics}}, are available.
Arb has competitive performance compared with MPFR, despite being based on interval arithmetic, and is therefore of interest where a large number of tests need to be computed.
In \texttt{v0.3} of the test bench we have opted for \texttt{BigFloat} as it is a natively supported Julia type, and is backed by MPFR which provides correctly rounded operations to the specified precision.

\subsection{Floating-point arithmetic}\label{sec:fp}

\begin{table*}[t]
  \centering
  \caption{Features of widely used floating-point formats. Here $\smin$ and $\fmin$ are the smallest positive subnormal and normal values, respectively, $\fmax$ is the largest representable positive value, $\emin$ and $\emax$ denote the exponent range, and $p$ is the precision. The right-most column shows how many finite numbers are in each of the formats which determines the maximum number of tests that a test bench may need to run for one function. The bottom three formats are available in the Julia Base whilst bfloat16 is not a built-in type.}
  \begin{tabular}{lrlcllll}
    \toprule
    Type & $\emin$ & $\emax$ & $p$ & $\smin$ & $\fmin$ & $\fmax$ & Finite numbers \\
    \midrule
    bfloat16~\citep{inte18}    & -126 & 127 & 8 & $2^{-133}$ & $2^{-126}$ & $\sim 3.39 \times 10^{38}$ & $2^{16}-2^{8}=65280$ \\
    binary16~\citep{ieee19}    & -14 & 15 & 11 & $2^{-24}$ & $2^{-14}$ & $65504$ & $2^{16}-2^{11}=63488$ \\
    binary32~\citep{ieee19}  & -126 & 127 & 24 & $2^{-149}$ & $2^{-126}$ & $\sim 3.4 \times 10^{38}$ & $2^{32}-2^{24}=4278190080$ \\
    binary64~\citep{ieee19}  & -1022 & 1023 & 53 & $2^{-1074}$ & $2^{-1022}$ & $\sim 1.79 \times 10^{308}$  & $2^{64}-2^{53}\approx 1.8\times 10^{19}$ \\
    \bottomrule
  \end{tabular}
  \label{table:fp-formats}
\end{table*}

A nonzero binary floating-point number $x$ has the form $(-1)^s \times m \times 2^{e-p+1}$, where $s$ is the sign bit, $p$ is the precision, $m \in [0, \; 2^p-1]$ is the integer significand, and $e \in [e_{\min}, \; e_{\max}]$ is the integer exponent with $e_{\min}=1 - e_{\max}$~\cite{ieee19}.
In order for $x$ to have a unique representation, the number system is \textit{normalized} so that the most significant bit of $m$ is set to~1 if $|x| \ge 2^{e_{\min}}$.
Therefore, all floating-point numbers with $m \geq 2^{p-1}$ are normalized.
Numbers below the smallest normalized number $2^{e_{\min}}$ in absolute value are called \textit{subnormal numbers}, and are such that $e=e_{\min}$ and $0 \leq m < 2^{p-1}$.
A finite set of floating-point numbers is denoted by $\F$.
We denote the smallest positive subnormal, the smallest positive normal, and the largest positive normal numbers using $\smin$, $\fmin$, and $\fmax$ respectively, with the specific format to which these symbols refer to always clear from the context.
Table~\ref{table:fp-formats} lists the main formats and their features, including \texttt{bfloat16} used for the example above and which is widely in use for machine learning applications, but not listed in the IEEE 754 standard and not yet available in Julia Base.

The latest revision of the IEEE 754 standard~\citep{ieee19} defines six rounding modes.
Four rounding modes are required for a binary \fp\ arithmetic to be
compliant: round to nearest with ties to even ($\rn$), round toward
positive (or toward $+\infty$, or up, $\ru$), round toward negative (or toward
$-\infty$, or down, $\rd$), and round toward zero ($\rz$).
These are described formally in Table~\ref{table:rounding-modes}.
Julia supports these rounding modes\footnote{\url{https://docs.julialang.org/en/v1/base/math/\#Base.Rounding.RoundingMode}} in floating-point arithmetic.

\begin{table*}
\centering
\caption{A subset of the rounding modes defined in the 2019 revision of the IEEE 754 standard
  \citep{ieee19} and available in Julia.}
\label{table:round}
\begin{tabular}{ll}
  \toprule
  Round mode & Description \\
  \midrule
  To nearest ($\rn$)
             & $|\rn(x)-x|=\min\{|x'-x| : x' \in \F\}$, ties to even significand \\
  Toward positive ($\ru$) & $\ru(x) = \min\{x' \in \F : x \le x'\}$ \\
  Toward negative ($\rd$) & $\rd(x) = \max\{x' \in \F : x \ge x'\}$ \\
  Toward zero ($\rz$) & $\rz(x) = \sign(x) \cdot \max\{ x' \in \F : |x| \ge |x'|\}$ \\
  \bottomrule
\end{tabular}
\label{table:rounding-modes}
\end{table*}

\subsubsection{Unit of least precision (ULP)}

Informally, given $x \in \R$, $\ulp{x}$ is the difference between the two floating-point numbers nearest to $x$ in some $\F$ if $x \notin \F$.
If $x \in \F$, then $\ulp{x}$ is the difference between $x$ and the floating-point number that follows it towards $\sign(x)\cdot \infty$.
Formally~\citep[Sec.~2.1]{bhmz25},
\begin{align}\label{eq:ulp}
  \ulp{x}=
  \begin{cases}
    2^{\max(\emin, \lfloor \log_2|x| \rfloor)-p+1} \; & \mathrm{if} \; x \ne 0, \\
    2^{\emin-p+1} & \mathrm{otherwise.}
  \end{cases}
\end{align}

Notice that $\ulp{} $ is defined on reals and implicitly has $p$ in its definition, the value of which is always clear from context in this paper based on the floating-point format at hand.

\begin{center}
\fbox{
  \parbox{0.9\textwidth}{In Section~\ref{sec:results} we report the \emph{worst-case errors} or the \emph{known worst-case errors} in units of ULP. For example, 2.5ULPs of error tells us that the approximation of a function $f$ is not farther away from $f$ than $2.5\ulp{f(x)}$ for any $x$ in the input domain of the approximation of $f$, where $f(x)$ is a reference solution. For binary64, the reported error is on a subset of the input domain because exhaustive testing is not feasible.}
}
\end{center}

\subsection{The state of mathematical functions in binary floating point}

IEEE 754-2019~\citep[Sec.~9.2]{ieee19} standard recommends providing a set of mathematical functions, and if implemented, requires them to return a correctly rounded result for each rounding mode, resulting in the worst case error no larger than 0.5 or 1ULP and the ties broken correctly for round-to-nearest ties-to-even.
The OpenCL C Language Specification~\citep{khro25} defines full and embedded profiles, and specifies the minimum accuracy in each profile, separately for each function.
For example, for half-precision exponential function the specification requires approximation errors to be no larger than 2ULPs in the full profile and no larger than 3ULPs in a different, embedded profile.

An implementation of C standard library from GNU\footnote{\url{https://sourceware.org/glibc/manual/2.42/html_mono/libc.html\#Errors-in-Math-Functions}} remarks ``[...]\emph{ many of the functions in the math library have errors. The math testsuite only flags results larger than 9ulp (or 16 for IBM long double format) as errors; although most of the implementations show errors smaller than the limit}''.

The  CORE-MATH library of Sibidanov,~Zimmermann,~and~Glondu~\citep{szg22} provides a set of fast correctly rounded univariate mathematical function implementations for binary32.
For binary64 functions and bivariate binary32 functions, CORE-MATH either
delivers a correctly rounded result, or prints an error message when an unknown
hardest-to-round point is found~\citep{hjz23}.

RLIBM\footnote{\url{https://people.cs.rutgers.edu/~sn349/rlibm/}} provides several mathematical functions correctly rounded in the binary32 and lower precision formats.
It generates polynomials such that they approximate the correctly rounded result rather than the exact result, which then can be rounded to some lower precision with an appropriate rounding mode, producing a correctly rounded result in that precision.
To avoid errors associated with double rounding, RLIBM approximates the correctly rounded result for round-to-odd~\cite{sygu05} rounding mode.

Gladman,~Innocente,~Mather~and~Zimmermann~\citep{gimz25} have performed testing of various mathematical libraries, including the aforementioned GNU libc, Intel MKL, Apple Math Library, LLVM libc, and several others.
The tests were performed on binary32 (exhaustively for univariate functions), binary64, ``long double'' (80-bit floating-point representation on x86 architectures), and binary128.
Bivariate functions in binary32 and the rest of the formats were tested by utilizing three recursive search strategies for covering the large input space.
A threshold $t$ is set and if a given search space has more floating-point values than $t$ it is split into two or more parts, searching each part by checking $t$ random inputs and recursively repeating until the search space is sufficiently covered.
Sub-intervals with the largest error, maximum average error, or the largest expected error are chosen in the recursive step.
The authors found the first strategy, the maximum error in ULPs, to be the most effective, but all three are utilized in parallel.
The highlight result of this work is that LLVM libc was found to have the maximum error of 0.5ULP in binary32 and binary64, with other libraries yielding errors between 0.5 and 1ULP or in some cases significantly higher.
The square root in all of the tested libraries and formats was found to have the maximum error of 0.5ULP and we have also obtained this outcome in Julia.
The paper provides exact input values~\citep{gimz25} that yield the highest errors and we use these in our testing alongside the search strategies.

Ozaki~\cite{ozak25} has undertaken the testing of the accuracy of functions in MATLAB, Octave, CUDA, and other environments.
The author used interval arithmetic in binary64 to obtain the intervals containing exact results, then used those to obtain the intervals containing ulp errors.
Maximum errors between $0.51$ULP and $0.88$ULP were determined for 16 functions in binary16 in MATLAB.
For binary32, maximum errors between $0.86$ULP and $1024$ULP were found across 27 functions.
In GNU Octave, Ozaki discovered errors between $0.5$ULP and $1024$ULPs, except sin and cos which are reported to have even larger maximum errors.
Another notable result from this work is the accuracy of binary32 functions in Numpy: between $0.51$ULP and $3.84$ULPs (natural logarithm being the most inaccurate).
Functions in binary64 were also tested, but instead of utilising a search strategy to cover the large input space as Gladman et al.~\cite{gimz25} did, binary32 inputs were converted to binary64.
For binary64, maximum known errors between $0.77$ULP and $18243$ULPs were demonstrated across 27 functions.

Brisebarre,~Hanrot,~Muller,~and~Zimmermann~\citep{bhmz25} argue that while correct rounding of elementary functions has for a long time been looked at as infeasible, currently it is becoming feasible to design the algorithms (which involves finding hardest-to-round points), evaluate them cheaply, and to prove correct rounding.
The article recommends the next revision of IEEE 754 to require a core set of binary16, binary32, and binary64 (univariate) correctly rounded functions.

As an aside, the importance of elementary functions on the applications side is demonstrated by the following work.
Glatard et al.~\cite{glfa15} reported reproducibility issues due to differences between versions of mathematical libraries, in performing analysis of neuroimages in single precision floating-point arithmetic (binary32).
The differences stem from mathematical functions not being correctly rounded and therefore having varying levels of error.
They have concluded that to avoid the impact of differences in mathematical libraries, use of binary32 functions should be avoided and instead replaced by binary64 or higher-precision arithmetic where the variations are smaller.

\section{Methods}

\subsection{Calculating the maximum ULP errors}
\label{sec:max_ulp_error}

Given a set of monotonically increasing input values $\{x_1, \cdots, x_N\} \eqqcolon I$ in some floating-point system $\F  \supset I$ generated by \emph{a search strategy}, a function approximation $\widehat{f}$ and a reference function $f$,
\begin{equation}
  \max_{x \in I}\err{x}=\err{x_H}, \; \text{for any } x_H \in \argmax_{x \in I} \err{x},
\end{equation}
where
\begin{equation}
  \argmax_{x \in I}\err{x} = \{x : \err{x} \ge \err{y}, \forall y \in I\}
\end{equation}
and
\begin{equation}\label{eq:err}
  \err{x}:=\frac{|\widehat{f}(x)-f(x)|}{\ulp{\rz(f(x))}}.
\end{equation}
Here $\err{x_H}$ is the maximum error, measured in ULPs, of $\widehat{f}$ over the input domain $[x_1, x_N] \ni x_H$.
If $\err{x_H} < 0.5$ then the function $\widehat{f}$, relative to the reference $f$, is correctly rounded~\cite{ieee19}, for the RN mode, in the input range $[x_1, x_N]$.
For the other rounding modes of Table~\ref{table:rounding-modes} the property can be relaxed to $\err{x_H} < 1$, and the direction of rounding has to be checked to be correct for each rounding mode.
See Muller~\cite[Sec.~2.6]{boje23} for details.

In \eqref{eq:err} we take $\rz(f(x))$ in order to round the reference value $f(x)$ to the target precision before calculating the size of ULP, which is precision dependent.
Round-to-zero is used in order to avoid rounding away from zero and landing on a power of two, which would increment the ULP unit by a factor of two compared with the ULP between the two floating-point numbers surrounding $f(x)$.

We have to decide how to calculate the error \eqref{eq:err} and what accuracy the reference solution $f(x)$ should be computed to.
First of all, $\widehat{f}(x) \in \F$ and $\ulp{\rz(|f(x)|)} \in \F$ since according to \eqref{eq:ulp} it is a power of two and no smaller than $\smin =2^{\emin-p+1}$ and no larger than $\fmax =2^{e_{max}}$;
to assure this we need $\rz(f(x)) \neq \infty$, in other words, no overflows can occur when a high precision approximation of the function is rounded to the working precision.
We can assure this is the case by choosing the input range of the function under test such that no overflows would occur in a given floating-point format (Section~\ref{sec:input_space}).

Since $\ulp{\rz(f(x))}$ is a power of two, the division in \eqref{eq:err} is exact and the main question then is how to compute the numerator.
There are two options:
\begin{enumerate}
\item Round $f(x)$ to $\F$ (the data format of the approximation $\widehat{f}$) and compute \eqref{eq:err} in $\F$, or
\item compute \eqref{eq:err} in some higher precision.
\end{enumerate}
Option 1 would be suitable if we were interested in testing only whether  $\widehat{f}(x)$ is correctly rounded but not reporting the error precisely: taking $\rn(f(x))$ we would introduce an error of up to 0.5ULP in the reference result and then we would look for the numerator to evaluate to zero and could report the maximum error of 0.5ULP without actually knowing where it is in $[0, 0.5]$ULP.
In general we would introduce an error of up to 0.5ULP for round-to-nearest and no more than 1ULP for the directed rounding modes, in the computation of \eqref{eq:err}.
With option 1 we would only be able to report the quantity \eqref{eq:err} in steps of $1$ULP.
It is worth noting that the errors of mathematical functions are not always measured relative to the exact result and a correctly rounded result is taken instead.
For example, an interim report of IEEE P3109~\cite{ieee25}, a floating-point standard for machine learning, defines $\kappa$-approximate implementations that in effect use $\circ(f(x))$ instead of $f(x)$ , where $\circ \in \{\rn, \rz, \ru, \rd\}$, as a reference solution.

In order to calculate the error relative to $f(x)$ instead of $\widehat{f}(x)$ and report it more precisely than in steps of $1$ulp, we have to use option 2.
For binary16, binary32, and binary64, we have setup MPFR (BigFloat) for approximating $f(x)$ in \eqref{eq:err} to precisions 63, 63, and 127, respectively.
These precisions are enough to compute elementary functions accurately enough to determine which side of the breakpoint the result is on the hardest-to-round cases~\cite[Sec.~10.5.3]{mbdj18}.
For reporting the error in Sec.~\ref{sec:results}, the error in one of the high MPFR precisions is rounded towards zero in our software.
Then, the errors in the tables of Sec.~\ref{sec:results} that are smaller than $0.5$ULP indicate correct rounding, and errors that are reported to be exactly $0.5$ULP indicate that the high-precision error was above $0.5$ULP.
The software can be modified to print out the worst-case errors in full precision if required.
Note that an exact error of $0.5$ULP is not possible because the considered functions cannot produce a midpoint between two floating-point number; this is due to the results such as Hermite-Lindemann's theorem as discussed in detail by Brisebarre~et~al.~\cite[Sec.~4.1]{bhmz25}.

To avoid changing the rounding mode of MPFR to $\rz$ as required for \eqref{eq:err}, which is a global state and requires synchronisation amongst Julia threads, we first compute $\ulp{\rn(f(x))}$.
Then, if $$|\rn(f(x))|>|f(x)| \wedge \rn(f(x)) \in \{2^k: k \in [\emin+1, \emax]\}$$ we halve it to obtain $\ulp{\rz(f(x))}$.

\subsection{Input space search strategies}

Here we describe methods for forming the subset $I \subset \F$ of test inputs, defined in Section~\ref{sec:max_ulp_error}.

In \texttt{v0.3} of the test bench we have implemented five fixed-step-size\footnote{The size of the step between the floating-point numbers in the input space is constant in the number of floating-point members.} input space search strategies with self-explanatory user-facing names corresponding approximately to the expected run-time (first four) for testing a particular mathematical function for one floating-point format: \emph{seconds}, \emph{minutes}, \emph{hours}, \emph{days}, \emph{exhaustive}.
For example, if \emph{minutes} is specified for binary64, each function's $I$ will be constructed such that testing all inputs in $I$ would take approximately a minute for that function.
As a result, with the \emph{minutes} search strategy, since we are testing 24 univariate functions in \texttt{v0.3}, it should take about 24 minutes to run the tests for one floating-point format entry in \texttt{config.json} (see Section~\ref{sec:test-bench}).

Thereafter we will refer to some of the source files in our Julia code, such as \texttt{functions.jl}, \texttt{MathBenchmark.jl}, and \texttt{config.json}.
The source is available.\footnote{\url{https://github.com/north-numerical-computing/MathBenchmark.jl}}

Given an input domain in $\F$ of a function $f$ for a particular floating-point format, $I_f$, form $I \subset I_f$ according to one of the strategies below.
Take $N$ members of $I_f$ to compile $I = \{x_1,\dots,x_N$\}. $N$ is set-up automatically for the testing of one mathematical function to take approximately the specified duration by the user, by measuring the average time taken $t$ (ns) for testing the accuracy of the function at one input point (which includes calling the corresponding reference implementation and computing \eqref{eq:err}) and setting
\begin{itemize}
\item Seconds: $N= \lfloor \frac{10^9}{t} \rfloor$.
\item Minutes: $ N= \lfloor \frac{60\times 10^9}{t} \rfloor$.
\item Hours: $N= \lfloor \frac{3600\times 10^9}{t} \rfloor$.
\item Days: $N= \lfloor \frac{24 \times 3600\times 10^9}{t} \rfloor$.
\item Exhaustive: $I=I_f$, $N=\mathrm{length}_{\F}(I_f)$, where $\mathrm{length}_{\F}(I_f)$ denotes the number of floating-point members of format $\F$ in function's $f$ input domain $I_f$.
\end{itemize}
In \texttt{v0.3} the distribution of the work to threads is done as follows.
For each function listed in \texttt{functions.jl} for a particular data format we first run a sample test on $10^5$ input arguments with all the available threads, measure the time taken, and compute the average duration of testing one input.
Then, based on the search strategy duration for each function, seconds, minutes, hours, or days, we calculate how many inputs we can test in that duration, $N$, using the formulas above.
Having $N$ we take the fixed-step size by dividing the number of floating-point members in a sub-range $i$ of $I_f$ by $N$: $\lceil \mathrm{length}_{\F}(I_f^i)/N \rceil$.

Following that, the function's input range $I_f$, specified in \texttt{functions.jl} for each of the three floating-point formats, is split up into approximately equal chunks to be scheduled across the number of threads available.
Each thread is then spawned to do approximately $N$ tests across several subsets of $I_f$, by assigning the sub-ranges of the input space to threads when they become available.
The total number of tests is $N\times P$ where $P$ is the number of cores on the testing system ($P=168$ in our case).
In general $N$ will be smaller than the number of members in a thread's subset of $I_f$ for binary32, unless a very large number of cores is used or a long run-time is specified, and almost always smaller for binary64.

As an alternative, the search strategy option of the configuration file (see Sec.~\ref{sec:test-bench}) can take an integer value, which will set $N$, the number of tests to run per thread, to that exact quantity.

\subsection{Determining input ranges of functions in different formats}
\label{sec:input_space}

\begin{table}[t!]
  \centering
  \caption{Exact input domains of mathematical functions over the set of binary16 floating-point values for which $|f(x)| \leq \fmax$ in binary16.}
  \label{table:bin16_input_dom}
  \begin{tabular}{rll}
    \toprule
    Function & Input domain $I_f$ \\
    \midrule
      acos &  $[-1.0, 1.0]$ \\
      acosh &  $[1.0, \fmax]$ \\
      asin &  $[-1.0, 1.0]$ \\
      asinh &  $[-\fmax, \fmax]$ \\
      atan &  $[-\fmax, \fmax]$ \\
      atanh &  $[\mathrm{nextfloat}(-1.0), \mathrm{prevfloat}(1.0)]$ \\
      cbrt &  $[-\fmax, \fmax]$ \\
      cos & $[-\fmax, \fmax]$ \\
      cosh & $[-11.78125, 11.78125]$ \\
      exp & $[-16.625, 11.0859375]$ \\
      exp10 &  $[-7.22265625, 4.8125]$ \\
      exp2 &  $[-24, 15.9921875]$ \\
      log & $[\smin, \fmax]$ \\
      log10 &  $[\smin, \fmax]$ \\
      log1p &  $[\mathrm{nextfloat}(-1)), \fmax]$ \\
      log2 &  $[\smin, \fmax]$ \\
      sin & $[-\fmax, \fmax]$ \\
      sinh &  $[-11.78125, 11.78125]$ \\
      sqrt &  $[0.0, \fmax]$ \\
      tan & $[-\fmax, \fmax]$ \\
      tanh &  $[-4.50390625, 4.50390625]$ \\
      cospi &  $[-\fmax, \fmax]$ \\
      sinpi &  $[-\fmax, \fmax]$ \\
      tanpi &  $[-\fmax, \fmax]$ \\
    \bottomrule
  \end{tabular}
\end{table}

\begin{table}[h!]
  \centering
  \caption{Exact input domains of mathematical functions over the set of binary32 floating-point values for which $|f(x)| \leq \fmax$ in binary32.}
  \label{table:bin32_input_dom}
  \begin{tabular}{rll}
    \toprule
    Function & Input domain $I_f$ \\
    \midrule
      acos &  $[-1.0, 1.0]$ \\
      acosh &  $[1.0, \fmax]$ \\
      asin &  $[-1.0, 1.0]$ \\
      asinh &  $[-\fmax, \fmax]$ \\
      atan &  $[-\fmax, \fmax]$ \\
      atanh &  $[\mathrm{nextfloat}(-1.0), \mathrm{prevfloat}(1.0)]$ \\
      cbrt &  $[-\fmax, \fmax]$ \\
      cos & $[-\fmax, \fmax]$ \\
      cosh & $[-89.415985107421875, \; 89.415985107421875]$ \\
      exp & $[-103.27892303466796875, \; 88.72283172607421875]$ \\
      exp10 &  $[-44.853466033935546875, \; 38.53183746337890625]$ \\
      exp2 &  $[-149, \; 127.99999237060546875]$ \\
      log & $[\smin, \fmax]$ \\
      log10 &  $[\smin, \fmax]$ \\
      log1p &  $[\mathrm{nextfloat}(-1)), \fmax]$ \\
      log2 &  $[\smin, \fmax]$ \\
      sin & $[-\fmax, \fmax]$ \\
      sinh &  $[-89.415985107421875, \; 89.415985107421875]$ \\
      sqrt &  $[0.0, \fmax]$ \\
      tan & $[-\fmax, \fmax]$ \\
      tanh &  $[-9.01091289520263671875, \; 9.01091289520263671875]$ \\
      cospi &  $[-\fmax, \fmax]$ \\
      sinpi &  $[-\fmax, \fmax]$ \\
      tanpi &  $[-\fmax, \fmax]$ \\
    \bottomrule
  \end{tabular}
\end{table}

The binary16 format can represent $63488$ finite floating-point numbers (Table~\ref{table:fp-formats}), therefore exhaustive testing of univariate functions is well within reach.
For the bivariate functions such as $x^y$ there are up to $63488^2=4030726144$ cases to check, which may also be within reach to cover exhaustively.

The binary32 format can represent $4278190080$ finite floating-point numbers.
While exhaustive testing of univariate functions is feasible, the bivariate functions will in practice be out of reach, since they may require running up to $4278190080^2\approx 1.8 \times 10^{19}$ tests.

The binary64 format can represent approximately $1.8 \times 10^{19}$ finite values and therefore testing univariate functions exhaustively may be out of reach, except perhaps those functions that have narrow input ranges.
Testing bivariate functions exhaustively is not possible at present, requiring approximately $10^{38}$ tests to be run.

Instead of trying all finite numbers in a format it may be beneficial to derive the input ranges of each of the functions in which the function is defined, does not overflow if correctly rounded to any of the standard rounding modes, and where it does not converge to a constant output value, such as zero when raising $e$ to large negative power.
In \texttt{functions.jl} we have hardcoded the exact endpoints of input ranges for all of the tested functions, for each of the formats.
The binary16 ranges are given in Table~\ref{table:bin16_input_dom} and the binary32 function ranges in Table~\ref{table:bin32_input_dom}.
In some cases the input range is the full range of representable values in a given floating-point format or a simple subset of it, such as the positive or the negative subset.
In some cases the input ranges are more complicated.
For example, the exponential
function overflows or underflows with the inputs of relatively low magnitude.
We can take $e^{\log(\fmax)}=\fmax$, but in general $\log(\fmax)\notin \F$.
However, taking $\rd(\log(\fmax))$ would assure that $e^{\rd(\log(\fmax))} \leq \fmax$.
For the bottom end of the range, we want to find the largest magnitude negative input value such that the exponential function does not underflow to zero.
Taking $\ru(\log(\smin))$ will assure that $e^{\ru(\log(\smin))}\geq \smin$.
For example, for binary16 this yields the input range of
$$[\ru(\log(2^{-24}))=-16.625, \; \rd(\log(65504))= 11.0859375].$$
It is worth noting, however, that this does not assure that the approximation of the mathematical function in Julia will not yield infinity.
When such cases occur, we do not take them into account in reporting the maximum error and report their occurrence separately---indeed if the approximation returns an infinity, there is no way to calculate the error since we do not know how far away from the reference result the approximation was before it was mapped to an infinity.
In the future iterations of the software, we will add explicit checks for the correctly returned infinities, as well as check for other aspects of the IEEE 754 conformance.

The most complicated input domain is that of $\tanh=\frac{e^x-e^{-x}}{e^{x}+e^{-x}}$: this function approaches $1$ for $x\to \infty$ and $-1$ for $x\to -\infty$, so we are interested in finding the maximum magnitude inputs for which $\rn(\tanh(x))<1$.
We can derive a formula by checking for which input $x$ do we have $\frac{e^x-e^{-x}}{e^{x}+e^{-x}}=1-u/2$.
  Here $1-u/2$ is a midway point between $1$ and the preceding floating-point member in some $\F$ with $u=2^{-p}$.
  Once we have $x$ such that the value of the function is on the midway point, we can round $x$ towards zero to assure that the output from the function does not round up to $1$, or take a floating-point member preceding $x$ if $x \in \F$.
  We have
  $$
  \frac{e^x-e^{-x}}{e^{x}+e^{-x}}=1-u/2,
  $$
  $$
  e^x-e^{-x}=(1-u/2)(e^x+e^{-x}),
  $$
  $$
  e^{2x}=\frac{4-u}{u},
  $$
  then
  $$
  x = 1/2\log\Bigl( \frac{4-u}{u} \Bigr).
  $$
  For example, for binary16 this gives $\rz\Bigl(1/2\log\Bigl( \frac{4-2^{-11}}{2^{-11}} \Bigr)\Bigr)=4.50390625$, and therefore the input range of the hyperbolic tangent in this format is
  $$[-4.50390625, 4.50390625],$$ because one step further to either side converges.

  We have used similar reasoning to derive all the input ranges for binary16 shown in Table~\ref{table:fp-formats}, binary32 shown in Table~\ref{table:bin32_input_dom} and binary64 shown in \texttt{functions.jl}.
  For some functions, such as tanpi, values close to periodic vertical asymptotes may be in $I_f$ where the function tends to infinity, while the surrounding members in $I_f$ would not cause it.
  For these type of functions we report the number of infinities that appeared to inform the user, and we skip the tests that produce infinities but continue testing other members of $I_f$.
  In later versions of the test bench we plan to investigate how to detect the inputs that are either exactly on or very close to the vertical asymptotes of a function and skip them without computing the function.

\subsection{Utilizing known hardest-to-round points}

Vincent Lef\`{e}vre's website\footnote{\url{http://perso.ens-lyon.fr/jean-michel.muller/Intro-to-TMD.htm}} contains hardest-to-round points for various mathematical functions.
We have incorporated functionality in \texttt{functions.jl} to include special inputs for each function that should be tested in addition to the set of test inputs $I$ formed by the chosen search strategy.
We have included some inputs that caused largest errors in the study of Gladman et al.~\cite{gimz25}.
Since the authors mention including Vincent Lef\`{e}vre's reported points, they should probably appear in those tables, albeit not necessarily if the function is for some reason implemented to perform worst not on its hardest to round points.
In \texttt{v0.3} of the Julia test bench we have incorporated binary32 inputs causing maximum errors in AMD LibM 5.1 and Newlib 4.5.0 as reported in the source code of Gladman et al.~\cite{gimz25}\footnote{\url{https://gitlab.inria.fr/zimmerma/math_accuracy/-/blob/master/latex/glibc242.tex?ref_type=heads}} and binary64 inputs causing maximum errors in GNU libc 2.41 and IML 2025.0.0 as reported in \cite[Table~4]{gimz25}.
We have also included a few worst case inputs for each of the binary64 functions from the CORE-MATH project's set of \texttt{.wc} files available at \url{https://gitlab.inria.fr/core-math/core-math/-/tree/master/src/binary64} which also include Lef\`{e}vre's cases.

\section{Results}\label{sec:results}

\subsection{The test bench}
\label{sec:test-bench}

Our test bench for measuring the accuracy of Julia's mathematical function is available on GitHub.

Below is an example JSON setup file for testing mathematical functions in binary16, binary32 and binary64 formats, using ``exhaustive'' and ``hours'' search strategies.
The test bench reads the test descriptor file and performs the specified tests of functions that are listed in \texttt{functions.jl} for each format, with the specified rounding mode, fast math mode, and the search strategy.
In \texttt{v0.3} the rounding mode is a placeholder for future functionality: as far as we are aware Julia does not support mathematical functions rounded with four rounding modes in Table~\ref{table:round}, presumably because the functions at present are not correctly rounded.
It is worth noting that a wrapper for CRlibm exists\footnote{\url{https://github.com/JuliaIntervals/CRlibm.jl}} which does provide access to correctly rounded functions for the standard rounding modes, in Julia.

The test bench produces a number of output files documenting the results, named after the test name; for example, for the first test defined in the JSON setup file below, it produces

\texttt{outputs/test-binary16RN-exhaustive-nofastmath.txt}.

\begin{lstlisting}[language=json]
{
  "test-binary16RN-exhaustive-nofastmath" : {
    "format" : "binary16",
    "rounding" : "RN",
    "fastmath" : 0,
    "search" : "exhaustive"
  },
  "test-binary32RN-exhaustive-nofastmath" : {
    "format" : "binary32",
    "rounding" : "RN",
    "fastmath" : 0,
    "search" : "exhaustive"
  },
  "test-binary64RN-hours-nofastmath" : {
    "format" : "binary64",
    "rounding" : "RN",
    "fastmath" : 0,
    "search" : "hours"
  }
}
\end{lstlisting}

In the future iterations of the software we plan to include tests for bivariate functions available in Julia Base.
For those cases that are not feasible to test exhaustively, such as $x^y$, the search strategy that the user specifies will apply across both inputs such that the run-time duration is still preserved.
For example, once $N$ is calculated, a grid-like choice of inputs $x$ and $y$ can be done such that overall $N$ tests are performed.

The workflow is simple: specify the above configuration in \texttt{config.json} and then run \texttt{julia -t <T> run.jl} in order to invoke the tests outlined in the configuration.
Here \texttt{<T>} is the number of threads to be used for testing, which can be \texttt{auto} or an integer value.

\subsection{System specification}

For running the tests specified in the JSON setup in Section~\ref{sec:test-bench} we have utilized one node of Aire high-performance computing platform installed at the University of Leeds.\footnote{\url{https://arc.leeds.ac.uk/platforms/aire/}}
Aire's nodes contain two AMD EPYC 9634 84-core CPUs running at 2.2GHz, overall providing 168 cores.
We have used Julia 1.12.7 which depends on OpenLibm 0.8.7.

\subsection{Accuracy of functions in Julia}

\subsubsection{Functions approximated for the binary16 format}

Tables~\ref{table:bin16_res}~and~\ref{table:bin16_res_hex} provide the maximum errors for 24 univariate functions for the binary16 format.
These have been tested exhaustively by covering every binary16 floating-point number in each of the input domains specified in Table~\ref{table:bin16_input_dom}.
All functions have errors close to $0.5$ULP.
Functions sqrt, sinh, asin, cospi, sinpi, cbrt, atanh, log2, and tanh are correctly rounded with maximum errors lower than $0.5$ULPs.
We also ran these tests in Julia's @fastmath mode and the results are available with the software on the GitHub repository.

In terms of comparison, Gladman et al.~\cite{gimz25} did not undertake any binary16 function testing, but Ozaki~\cite{ozak25} did in MATLAB and CUDA, so we can compare Julia's binary16 function accuracy to those results.
In terms of MATLAB~\cite[Table~4]{ozak25}, all tested functions are close to $0.5$ULP, as are those in Julia; the only outlier is the exponential function which has an error of $0.88$ULP in MATLAB---in Julia it has an error of $0.50003$ULP.
In CUDA, all the tested functions~\cite[Table~9]{ozak25} are correctly rounded except sin which has an error of $0.96$ULP---in Julia we found $0.50004$ULP.

\pgfplotstableread[col sep=space]{./data/test-binary16RN-exhaustive-nofastmath.txt}\half
\pgfplotstableread[col sep=space]{./data/HEX_test-binary16RN-exhaustive-nofastmath.txt}\halfhex
\pgfplotstableread[col sep=space]{./data/test-binary32RN-exhaustive-nofastmath.txt}\single
\pgfplotstableread[col sep=space]{./data/HEX_test-binary32RN-exhaustive-nofastmath.txt}\singlehex
\pgfplotstableread[col sep=space]{./data/test-binary64RN-hours-nofastmath.txt}\double
\pgfplotstableread[col sep=space]{./data/HEX_test-binary64RN-hours-nofastmath.txt}\doublehex

\begin{table}[ht]
  \centering
  \caption{Accuracy of Julia's mathematical library in binary16 with ``exhaustive'' input space search strategy. The ``Input'' and ``Output'' columns provide the input and output points at which the maximum error was measured. The MPFR column provides the value that MPFR computes for that function for the exact same input. The column ``Tests'' provides the number of different binary16 inputs that the functions have been tested on.}
  \label{table:bin16_res}
  \pgfplotstabletypeset[
  columns={Function, ULPs, Input, Output, MPFR, Tests},
    sort=true,
  sort cmp={string < },
    columns/Function/.style={string type, column type=p{0.1\textwidth}},
    columns/ULPs/.style={precision=5, column type=p{0.1\textwidth}},
    columns/Input/.style={precision=10, column type=>{\raggedleft\arraybackslash}p{0.17\textwidth}},
    columns/Output/.style={precision=10, column type=>{\raggedleft\arraybackslash}p{0.19\textwidth}},
    columns/MPFR/.style={precision=10, column type=>{\raggedleft\arraybackslash}p{0.19\textwidth}},
    columns/Tests/.style={column type=>{\raggedleft\arraybackslash}p{0.1\textwidth}},
    every head row/.style={before row=\toprule, after row=\midrule},
    every last row/.style={after row=\bottomrule}
  ]{\half}
\end{table}

\captionsetup{width=\textwidth}
\begin{table}[t!]
  \centering
  \caption{
    Errors of Julia's mathematical library in binary16 with ``exhaustive'' input space search strategy. The ``Input'' and ``Output'' columns provide the hexadecimal representation of the encodings of input and output points at which the maximum error was measured.}
  \label{table:bin16_res_hex}
  \pgfplotstabletypeset[
  columns={Function,ULPs,Input,Output},
  sort=true,
    sort cmp={string < },
    columns/Function/.style={string type, column type=p{0.1\textwidth}},
    columns/ULPs/.style={precision=5, column type=p{0.1\textwidth}},
    columns/Input/.style={string type, column type=>{\raggedleft\arraybackslash}p{0.1\textwidth},
          postproc cell content/.append style={
        /pgfplots/table/@cell content=\texttt{##1}
      }},
  columns/Output/.style={string type, column type=>{\raggedleft\arraybackslash}p{0.1\textwidth},
        postproc cell content/.append style={
        /pgfplots/table/@cell content=\texttt{##1}
      }},
    columns/Tests/.style={column type=>{\raggedleft\arraybackslash}p{0.1\textwidth}},
    every head row/.style={before row=\toprule, after row=\midrule},
    every last row/.style={after row=\bottomrule}
]{\halfhex}
\end{table}

\subsubsection{Functions approximated for the binary32 format}

Table~\ref{table:bin32_res} contains the maximum errors in 24 Julia's univariate mathematical functions for the binary32 floating-point data type.
These are also obtained by testing every binary32 number in each function's input domain shown exactly in Table~\ref{table:bin32_input_dom}.
Table~\ref{table:bin32_res_hex} contains the binary strings in hexadecimal encoding of the inputs and outputs that cause maximum errors in each of the functions.

We can compare these binary32 results to the accuracy of 13 mathematical libraries reported by Gladman et al.~\cite[Table~1]{gimz25}.
First of all, the highest errors in Julia are exhibited by sinh and cosh functions: $\sim 2.4$ULPs.
Gladman~et~al.~\cite[Table~1]{gimz25} show that newlib and CUDA have larger errors than that, but the rest of the 13 libraries are more accurate than Julia.
For cosh, only Newlib has a slightly higher error than Julia.
Functions sin, cos, sinpi, cospi, and tanpi exhibit maximum errors very close to $0.5$ULP in Julia; some of these implementations are more accurate than, for example, GNU libc, IML, and AMD implementations~\cite[Table~1]{gimz25}.
Notably, Julia's log function has $0.55$ULPs of error, which is more accurate than 11 of the implementations reported by Gladman~et~al.~\cite[Table~1]{gimz25}.
Exp function, on the other hand, is more accurate than only 5 out of the 13 implementations.
Finally, only sqrt and cbrt are correctly rounded, whilst cospi, sinpi, tanpi, cos, and sin are very close to $0.5$ULPs.
Other functions have been measured to have the worst case error of between $0.55$ULP and $2.42$ULPs.

\begin{table}[ht]
  \centering
  \caption{Accuracy of Julia's mathematical library in binary32 with ``exhaustive'' input space search strategy. The ``Input'' and ``Output'' columns provide the input and output points at which the maximum error was measured. The column ``Tests'' provides the number of different binary32 inputs that the functions have been tested on.}
  \label{table:bin32_res}
  \pgfplotstabletypeset[
  columns={Function,ULPs,Input,Output,Tests},
  sort=true,
  sort cmp={string < },
    columns/Function/.style={string type, column type=p{0.1\textwidth}},
    columns/ULPs/.style={precision=10, column type=p{0.1\textwidth}},
    columns/Input/.style={precision=5, column type=>{\raggedleft\arraybackslash}p{0.19\textwidth}},
    columns/Output/.style={precision=5, column type=>{\raggedleft\arraybackslash}p{0.19\textwidth}},
    columns/MPFR/.style={precision=5, column type=>{\raggedleft\arraybackslash}p{0.19\textwidth}},
    columns/Tests/.style={precision=10, fixed, column type=>{\raggedleft\arraybackslash}p{0.18\textwidth}},
    every head row/.style={before row=\toprule, after row=\midrule},
    every last row/.style={after row=\bottomrule}
  ]{\single}
\end{table}

\begin{table}[ht]
  \centering
  \caption{Errors of Julia's mathematical library in binary32 with ``exhaustive'' input space search strategy. The ``Input'' and ``Output'' columns provide the hexadecimal representation of the encodings of input and output points at which the maximum error was measured.}
  \label{table:bin32_res_hex}
  \pgfplotstabletypeset[
  columns={Function,ULPs,Input,Output},
  sort=true,
    sort cmp={string < },
    columns/Function/.style={string type, column type=p{0.1\textwidth}},
    columns/ULPs/.style={precision=5, column type=p{0.1\textwidth}},
    columns/Input/.style={string type, column type=p{0.17\textwidth},
      postproc cell content/.append style={
        /pgfplots/table/@cell content=\texttt{##1}
      }},
    columns/Output/.style={string type, column type=p{0.13\textwidth},
    postproc cell content/.append style={
        /pgfplots/table/@cell content=\texttt{##1}
      }},
    columns/Tests/.style={precision=10, fixed,column type=>{\raggedleft\arraybackslash}p{0.17\textwidth}},
    every head row/.style={before row=\toprule, after row=\midrule},
    every last row/.style={after row=\bottomrule}
]{\singlehex}
\end{table}

\subsubsection{Functions approximated for the binary64}

Table~\ref{table:bin64_res} contains the maximum known errors in 24 Julia's univariate mathematical functions for the binary64 floating-point data type.
These are obtained by testing billions of the finite binary64 numbers by running each function's testing for approximately 1 hour on a 168-core CPU; note that it is still a very small proportion of the finite numbers representable in binary64 and this is a lower bound on the maximum error.
Table~\ref{table:bin64_res_hex} contains the binary strings in hexadecimal encoding of the inputs and outputs that cause maximum errors in each of the functions.

These results can be compared with the lower bounds of the maximum errors of the 13 libraries tested by Gladman~et~al.~\cite[Table~3]{gimz25}.
We note that their search strategy is different from ours, which means the input set and the proportion of the whole input space covered is different from ours.
One notable observation is that all functions, except sqrt, are not correctly rounded.
This is in line with Gladman~et~al.~\cite[Table~3]{gimz25} who found that only LLVM has correctly rounded binary64 functions, other than sqrt, assuming the lower bound discovered by testing a small sample of inputs is in fact the upper bound.

\begin{table}[ht]
  \centering
  \caption{Accuracy of Julia's mathematical library in binary64 with ``hours'' input space search strategy. The ``Input'' and ``Output'' columns provide the input and output points at which the maximum error was measured. The column ``Tests'' provides the number of different binary64 numbers that the functions have been tested on.}
  \label{table:bin64_res}
  \pgfplotstabletypeset[
  columns={Function,ULPs,Input,Output,Tests},
  sort=true,
    sort cmp={string < },
      columns/Function/.style={string type, column type=p{0.1\textwidth}},
    columns/ULPs/.style={precision=10, column type=p{0.1\textwidth}},
    columns/Input/.style={precision=5, column type=>{\raggedleft\arraybackslash}p{0.19\textwidth}},
    columns/Output/.style={precision=5, column type=>{\raggedleft\arraybackslash}p{0.19\textwidth}},
    columns/MPFR/.style={precision=5, column type=>{\raggedleft\arraybackslash}p{0.19\textwidth}},
    columns/Tests/.style={precision=10, fixed, column type=>{\raggedleft\arraybackslash}p{0.18\textwidth}},
    every head row/.style={before row=\toprule, after row=\midrule},
    every last row/.style={after row=\bottomrule}
  ]{\double}
\end{table}

\begin{table}[ht]
  \centering
  \caption{Errors of Julia's mathematical library in binary64 with ``hours'' input space search strategy. The ``Input'' and ``Output'' columns provide the hexadecimal representation of the encodings of input and output points at which the maximum error was measured.}
  \label{table:bin64_res_hex}
  \pgfplotstabletypeset[ columns={Function,ULPs,Input,Output},
  sort=true,
    sort cmp={string < },
    columns/Function/.style={string type, column type=p{0.1\textwidth}},
    columns/ULPs/.style={precision=5, column type=p{0.1\textwidth}},
    columns/Input/.style={string type, column type=p{0.26\textwidth},
      postproc cell content/.append style={
        /pgfplots/table/@cell content=\texttt{##1}
      }},
    columns/Output/.style={string type, column type=p{0.26\textwidth},
    postproc cell content/.append style={
        /pgfplots/table/@cell content=\texttt{##1}
      }},
    columns/Tests/.style={precision=10, fixed,column type=p{0.12\textwidth}},
    every head row/.style={before row=\toprule, after row=\midrule},
    every last row/.style={after row=\bottomrule}
]{\doublehex}
\end{table}

\section{Conclusion}

We have reported the worst-case errors of a subset of Julia's mathematical functions in binary16 and binary32 formats, and lower error bounds in binary64.
We have also presented the software with a simple configuration which will help the community to easily generate the presented accuracy results on new Julia versions in the future.
Our work has also helped detect a deadlock in Julia, related to the garbage collector and MPFR at high number of threads.\footnote{\url{https://github.com/JuliaLang/julia/issues/62753}}

The software in \texttt{v0.3} is available on GitHub at \url{https://github.com/north-numerical-computing/MathBenchmark.jl}.
In the next iteration we plan to expand it with the following features.
\begin{itemize}
\item Implement more sophisticated search strategies other than a fixed-size stepping, such as the ones employed by Gladman et al.~\cite{gimz25}.
\item Check a bigger proportion of binary64 inputs than we could cover here---this requires running the tests longer than one hour per function. We further plan to convert the software to an MPI-based model, allowing to deploy it over multiple nodes.
\item Implement testing of bivariate functions such as $x^y$.
\item Include all worst cases recorded in the \texttt{.wc} files at \url{https://gitlab.inria.fr/core-math/core-math/}.
\end{itemize}
As Julia is becoming more popular and as its base is being constantly developed, we believe our software will be of use in performing regression tests of its mathematical functions.
Running exhaustive binary16 and binary32 mathematical function accuracy tests is feasible as a regression test that is triggered upon changes to the implementations of the functions.
It will especially be useful if Julia developers start working on providing correctly rounded mathematical functions for several rounding modes of IEEE 754.

\section{Acknowledgements}

We are grateful to Paul Zimmermann for commenting on a draft of the paper and for discussions that led us to discover performance issues in our Julia code and subsequently fix those issues.
This work was undertaken on the Aire HPC system at the University of Leeds, UK and we thank John Hodrien for providing technical support on using the machine.
The first author was funded by the EPSRC grant ``\emph{Informing Future Numerical Standards by Determining Features of Non-Standard Mathematical Hardware}'' with the project reference 151: \url{https://gtr.ukri.org/projects?ref=UKRI151}.

\bibliographystyle{abbrvnat}
\bibliography{references}

\end{document}